\documentclass[aps,prl,reprint,twocolumn,superscriptaddress,floatfix,nofootinbib,longbibliography]{revtex4-2}
\usepackage{graphicx,amsmath,amsfonts,amssymb,xr}
\usepackage{epsfig, color, dsfont, upgreek, physics}
\usepackage{mathrsfs}
\usepackage{mathtools}
\usepackage{bbold}
\usepackage{comment}
\usepackage{braket}
\usepackage{lipsum}
\usepackage{soul}
\usepackage{bm}

\usepackage[bookmarks=true,colorlinks,linkcolor=OrangeRed,urlcolor=NavyBlue,citecolor=RoyalBlue]{hyperref}
\usepackage[capitalise]{cleveref}
\Crefname{figure}{Fig.}{Figs.}
\Crefname{section}{Sec.}{Secs.}
\usepackage[dvipsnames]{xcolor}
\usepackage[inline]{enumitem}
\setlist[enumerate]{label=(\roman*)}
\usepackage{cmds}
\graphicspath{{figures/}}
\usepackage{stackengine, scalerel}

\def\cmp{\mathbin{\ThisStyle{\ensurestackMath{\abovebaseline[-\dimexpr1.1pt+0.55\LMpt]{%
  \stackunder[-\dimexpr1pt+2.5\LMpt]{\color{darkgreen}\SavedStyle-}{%
  \color{red}\SavedStyle+}}}}}}
\usepackage{orcidlink}
\newcommand{\orcidgiovanni}{\orcidlink{0000-0002-9073-8978}}
\newcommand{\orcidjad}{\orcidlink{0000-0002-0659-7990}}
\newcommand{\orcidsimone}{\orcidlink{0009-0006-3594-8100}}
\newcommand{\LMU}{\affiliation{Department of Physics and Arnold Sommerfeld Center for Theoretical Physics (ASC), Ludwig Maximilian University of Munich, 80333 Munich, Germany}}
\newcommand{\MPQ}{\affiliation{Max Planck Institute of Quantum Optics, 85748 Garching, Germany}}
\newcommand{\MCQST}{\affiliation{Munich Center for Quantum Science and Technology (MCQST), 80799 Munich, Germany}}
\newcommand{\UNITO}{\affiliation{Dipartimento di Fisica, Università di Torino, I-10125 Torino, Italy}}
\newcommand{\KYUNGHEE}{\affiliation{Department of Physics, College of Science, Kyung Hee University, Seoul 02447, Republic of Korea}}
\begin{document}
\title{Real-Time String Dynamics in a \texorpdfstring{$2+1$}{2+1}D Non-Abelian Lattice Gauge Theory:\\ String Breaking, Glueball Formation, Baryon Blockade, and Tension Reduction}
\author{Giovanni Cataldi$^{\orcidgiovanni}$} \MPQ \MCQST
\author{Simone Orlando$^{\orcidsimone}$} \UNITO \MPQ \MCQST
\author{Jad C.~Halimeh$^{\orcidjad}$}\email{jad.halimeh@lmu.de}\LMU \MPQ \MCQST \KYUNGHEE
\date{\today}
\begin{abstract}
Understanding flux string dynamics can provide insight into quark confinement and hadronization. 
First-principles quantum and numerical simulations have mostly focused on toy-model Abelian lattice gauge theories (LGTs). With the advent of state-of-the-art quantum simulation experiments, it is important to bridge this gap and study string dynamics in non-Abelian LGTs beyond one spatial dimension. 
Using tensor network methods, we simulate the real-time string dynamics of a $2\!+\!1$D SU$(2)$ Yang--Mills LGT with dynamical matter. 
In the strong-coupling regime and at resonance, string breaking occurs through sharp Casimir reduction along with meson and baryon-antibaryon formation, a distinctively non-Abelian feature. 
At finite baryon density, we discover a \textit{baryon blockade} mechanism that delays string breaking.
Away from resonance, the magnetic term drives purely non-Abelian fluctuations: glueball loops and self-crossed strings that resolve two SU$(2)$ intertwiners with distinct dynamics. 
For higher-energy strings, we uncover representation-dependent tension-reduction resonances. Our findings serve as a guide for upcoming quantum simulators of non-Abelian LGTs.
\end{abstract}

\maketitle
\textbf{\textit{Introduction.---}}In non-Abelian gauge theories such as quantum chromodynamics (QCD), confinement manifests as color-electric flux strings between static charges; at large separations, these strings can break by producing dynamical matter, thereby flattening the heavy-quark potential and directly connecting to hadronization \cite{Weinberg1995QuantumTheoryFields, Ellis2003QCDColliderPhysics, Workman2022ReviewParticlePhysics}. 
Such phenomena lie beyond perturbation theory and inspired the development of lattice gauge theories (LGTs) \cite{Gattringer2009QuantumChromodynamicsLattice, Rothe2012LatticeGaugeTheories} in both Lagrangian \cite{Wilson1974ConfinementQuarks, Wilson1977QuarksStringsLattice} and Hamiltonian formulations \cite{Kogut1975HamiltonianFormulationWilsons}. 
Although Euclidean path integral Monte Carlo (MC) methods for LGTs \cite{Creutz1983MonteCarloComputations, Creutz1988LatticeGaugeTheory, Creutz1989LatticeGaugeTheories, Lynn2019QuantumMonteCarlo} have delivered benchmark results on confinement \cite{Creutz1979MonteCarloStudy, Creutz1980MonteCarloStudy, Berg1981SU2LatticeGauge} and static potential \cite{Trottier1999StringBreakingDynamical}, exploring finite-density regimes and real-time dynamics has presented significant challenges. 
There, MC methods suffer from sign problems that obstruct importance sampling and analytic continuation \cite{Kieu1994MonteCarloSimulations, Troyer2005ComputationalComplexityFundamental, Nagata2022FinitedensityLatticeQCD}. 

\begin{figure}[t!]
\includegraphics[width=1\columnwidth]{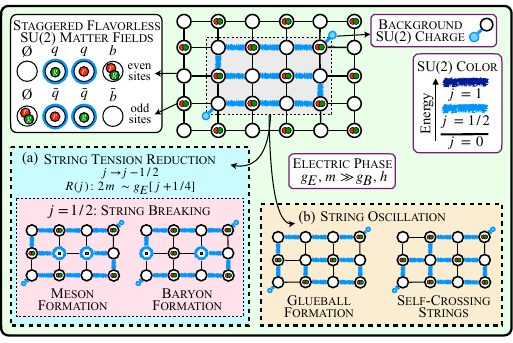}
\caption{\textbf{Real-time string dynamics in a $2\!+\!1$D SU$(2)$ LGT.} On a square lattice with staggered SU$(2)$ dynamical matter fields on sites and SU$(2)$ gauge fields on links, a generic string is created by connecting two static background SU$(2)$ charges within a (gray) patch. 
In a strong-coupling regime, the string undergoes two scenarios: (a) string breaking (and, more generally, tension reduction), where string links downgrade their Casimir $\spin\to\spin-\jonehalf$, allowing for meson and baryon creation; (b) string oscillations, where magnetic effects allow for glueballs and self-crossing strings. 
This scheme is for illustration purposes only; throughout this work, we consider a $4\!\times\!4$ string patch on an $8\!\times\!8$ lattice.
}
\label{fig1_scheme}
\end{figure}

These challenges have motivated alternative Hamiltonian approaches to LGTs that underlie the current major effort to quantum-simulate high-energy physics \cite{Byrnes2006SimulatingLatticeGauge, Dalmonte2016LatticeGaugeTheory, Zohar2015QuantumSimulationsLattice, Aidelsburger2021ColdAtomsMeet, Zohar2021QuantumSimulationLattice, Klco2022StandardModelPhysics, Bauer2023QuantumSimulationHighEnergy, Bauer2023QuantumSimulationFundamental,
DiMeglio2024QuantumComputingHighEnergy, Cheng2024EmergentGaugeTheory, Halimeh2022StabilizingGaugeTheories, Halimeh2023ColdatomQuantumSimulators, Cohen2021QuantumAlgorithmsTransport, Lee2025QuantumComputingEnergy, Turro2024ClassicalQuantumComputing, Bauer2025EfficientUseQuantum, Halimeh2025QuantumSimulationOutofequilibrium}, with recent years witnessing a large collection of impressive LGT quantum simulation experiments 
\cite{Martinez2016RealtimeDynamicsLattice, Klco2018QuantumclassicalComputationSchwinger,Gorg2019RealizationDensitydependentPeierls, Schweizer2019FloquetApproachZ2, Mil2020ScalableRealizationLocal, Yang2020ObservationGaugeInvariance, Wang2022ObservationEmergent$mathbbZ_2$, Su2023ObservationManybodyScarring, Zhou2022ThermalizationDynamicsGauge, Wang2023InterrelatedThermalizationQuantum, Zhang2025ObservationMicroscopicConfinement, Zhu2024ProbingFalseVacuum, Ciavarella2021TrailheadQuantumSimulation, Ciavarella2022PreparationSU3Lattice, Ciavarella2023QuantumSimulationLattice-1, Ciavarella2024QuantumSimulationSU3, 
Gustafson2024PrimitiveQuantumGates, Gustafson2024PrimitiveQuantumGates-1, Lamm2024BlockEncodingsDiscrete, Farrell2023PreparationsQuantumSimulations-1, Farrell2023PreparationsQuantumSimulations, 
Farrell2024ScalableCircuitsPreparing,
Farrell2024QuantumSimulationsHadron, Li2024SequencyHierarchyTruncation, Zemlevskiy2025ScalableQuantumSimulations, Lewis2019QubitModelU1, Atas2021SU2HadronsQuantum, ARahman2022SelfmitigatingTrotterCircuits, Atas2023SimulatingOnedimensionalQuantum, Mendicelli2023RealTimeEvolution, Kavaki2024SquarePlaquettesTriamond, Than2024PhaseDiagramQuantum, Angelides2025FirstorderPhaseTransition, Gyawali2025ObservationDisorderfreeLocalization, Cochran2025VisualizingDynamicsCharges, Gonzalez-Cuadra2025ObservationStringBreaking, Crippa2024AnalysisConfinementString, De2024ObservationStringbreakingDynamics, Liu2024StringBreakingMechanism, Alexandrou2025RealizingStringBreaking, 
Mildenberger2025Confinement$$mathbbZ_2$$Lattice, Schuhmacher2025ObservationHadronScattering, Davoudi2025QuantumComputationHadron, Cobos2025RealTimeDynamics2+1D, Saner2025RealTimeObservationAharonovBohm, Xiang2025RealtimeScatteringFreezeout, Wang2025ObservationInelasticMeson}.
In parallel to this body of experimental work, complementary and benchmarking studies of Hamiltonian-formulated LGTs have been performed using classical variational methods such as tensor networks (TNs) \cite{Schollwock2011DensitymatrixRenormalizationGroup, Banuls2019TensorNetworksTheir, Banuls2020ReviewNovelMethods, Montangero2018IntroductionTensorNetwork, Orus2019TensorNetworksComplex, Silvi2019TensorNetworksAnthology}. 
In the last two decades, they have provided fundamental contributions to static properties \cite{Banuls2013MassSpectrumSchwinger, Rico2014TensorNetworksLattice, Banuls2017EfficientBasisFormulation, Silvi2017FinitedensityPhaseDiagram, Magnifico2021LatticeQuantumElectrodynamics, Rigobello2023Hadrons1+1DHamiltonian, Cataldi2024Simulating2+1DSU2, Magnifico2025TensorNetworksLattice} and real-time dynamics of Abelian \cite{Pichler2016RealTimeDynamicsU1, Chanda2020ConfinementLackThermalization, Notarnicola2020RealtimedynamicsQuantumSimulation, Rigobello2021EntanglementGeneration$1+1mathrmD$, Magnifico2020RealTimeDynamics, VanDamme2022DynamicalQuantumPhase,Halimeh2022achievingquantum,Desaules2023ProminentQuantum, Osborne2023DisorderFreeLocalization$2+1$D,Osborne2024QuantumManyBodyScarring, VanDamme2023AnatomyDynamicalQuantumPhaseTransitions, Su2024ColdAtomParticleCollider, Belyansky2024HighEnergyCollisionQuarks, Jeyaretnam2025HilbertSpaceFragmentation, Xu2025StringBreakingDynamics, Tian2025RolePlaquetteTerm, Felser2020TwoDimensionalQuantumLinkLattice}
and non-Abelian \cite{Kuhn2015NonAbelianStringBreaking, Silvi2019TensorNetworkSimulation, Calajo2024DigitalQuantumSimulation, Calajo2025QuantumManybodyScarring-2,Cataldi2025DisorderFreeLocalizationFragmentation-1} LGTs.

In particular, recent TN simulations have uncovered new string dynamics phenomenology in $2+1$D Abelian LGTs, from the roughening transition
\cite{Marcantonio2025RougheningDynamicsElectric, Xu2025TensorNetworkStudyRoughening} to string breaking \cite{Borla2025StringBreaking$2+1$D, Xu2025StringBreakingDynamics, Tian2025RolePlaquetteTerm}, accompanied by various quantum simulation experiments in $1+1$D \cite{De2024ObservationStringbreakingDynamics, Liu2024StringBreakingMechanism, Alexandrou2025RealizingStringBreaking} and $2+1$D \cite{Cochran2025VisualizingDynamicsCharges, Gonzalez-Cuadra2025ObservationStringBreaking, Crippa2024AnalysisConfinementString}.
These results provide a concrete blueprint for nonequilibrium higher-dimensional LGT dynamics and motivate exploring this physics in non-Abelian models, where real-time string dynamics with dynamical matter, especially at finite density, has so far only been investigated in one spatial dimension \cite{Kuhn2015NonAbelianStringBreaking, Spitz2019SchwingerPairProduction}, and where quantum simulation proposals are emerging \cite{Atas2021SU2HadronsQuantum,Banerjee2013AtomicQuantumSimulation,KlcoSu2non-Abelian,Surace2024scalableabinitio,Halimeh2024SpinExchangeEnabled,depaciani2025quantumsimulationfermionicnonabelian}.

Our current work advances this frontier. 
Using tree tensor networks (TTNs), we simulate real-time string dynamics of a $2+1$D SU$(2)$ Yang--Mills LGT with dynamical matter at system sizes and timescales beyond current quantum-computing capabilities, while retaining \textit{ab initio} gauge invariance (also with background charges) and controlled access to finite-density sectors. 
We uncover purely non-Abelian signatures in two spatial dimensions, including baryon-antibaryon production at breaking, slowed-down breaking in finite density regimes, and intertwiner-resolved self-crossed strings away from breaking resonances.
In addition, for strings with nonminimal energy, we reveal representation-dependent tension-reduction phenomena. 
These results connect real-time SU$(2)$ string dynamics in $2\!+\!1$D to qualitative features relevant to $3\!+\!1$D QCD. They further provide concrete targets for near-term quantum simulators and next-generation TN methods for non-Abelian LGTs beyond one spatial dimension.

\textbf{\textit{Model.---}}To detect non-Abelian features of string breaking, we consider the $\mathrm{SU}(2)$ Yang--Mills LGT with flavorless dynamical matter \cite{Cataldi2024Simulating2+1DSU2, Calajo2024DigitalQuantumSimulation, Calajo2025QuantumManybodyScarring-2, Cataldi2025DisorderFreeLocalizationFragmentation-1} on a $2\!+\!1$D square lattice $\Lambda$ containing $\Nsites$ sites with indices $\vecsite\!=\!(\site[x],\site[y])$ and lattice spacing $\lspace$. 
Correspondingly, the lattice link connecting two neighboring sites $\vecsite$ and $\siteplus_{k}$ along the direction $k\in\qty{x,y}$ is denoted by $(\genlink_{k})$.
This model is described by the following Kogut--Susskind Hamiltonian \cite{Kogut1975HamiltonianFormulationWilsons} with \emph{staggered fermions} \cite{Susskind1977LatticeFermions}:
\begin{equation}
    \begin{split}
    \hat{H}=&-w\sum_{\vecsite,\alpha,\beta}
    \bigg[i\hpsi^{\dagger}_{\vecsite,\alpha}\hat{U}^{\alpha\beta}_{\vecsite,+ \latvec[x]}\hpsi_{\vecsite+\latvec[x],\beta}\\
    & +(-1)^{\site[x]+\site[y]}\hpsi^{\dagger}_{\vecsite,\alpha}\hat{U}^{\alpha\beta}_{\vecsite,+\latvec[y]}\hpsi_{\vecsite+\latvec[y],\beta}+\hc\bigg]\\
    &+ m\sum_{\vecsite,\alpha} (-1)^{\site[x]+\site[y]}\hpsi^{\dagger}_{\vecsite,\alpha}\hpsi_{\vecsite,\alpha}\\
    &+ \gauge[E]\sum_{\vecsite, k}\hat{E}^2_{\vecsite,+\latvec[k]}-\gauge[B]\sum_{\square}\qty(\plaq +\plaq*)
    \,. 
    \end{split}
    \label{eq_H}
\end{equation}
The first term describes the staggered hopping interaction, with $w\!=\!\frac{1}{2\lspace}$, between quark matter fields living on neighboring sites and $\mathrm{SU}(2)$ gauge fields on the links in between.
The quark field is represented as a staggered fermion $\hpsi_{\vecsite,\alpha}$, which satisfies the canonical anticommutation relations $\acomm*{\hpsi_{\vecsite,\alpha}}{\hpsi^{\dagger}_{\vecsite^{\prime},\beta}}\!=\! \delta_{\vecsite,\vecsite^{\prime}} \delta_{\alpha, \beta}$ and $\acomm*{\hpsi_{\vecsite,\alpha}}{\hpsi_{\vecsite^{\prime},\beta}}\!=\!0$
\cite{Susskind1977LatticeFermions}, where the $\alpha, \beta$ indices live in the fundamental $\mathrm{SU}(2)$ irreducible representation (irrep).
The local matter basis is given by:
\begin{math}
    \{
    \ket{\Omega},\,
    \hpsi^{\dagger}_{\rla}\ket{\Omega},\,
    \hpsi^{\dagger}_{\gla}\ket{\Omega},\,
    \hpsi^{\dagger}_{\rla} \hpsi^{\dagger}_{\gla}\ket{\Omega}
    \}
\end{math}, where $\ket{\Omega}$ represents an empty matter site, while $\qty{\rla,\gla}$ are shorthand notations for $\qty{\pm\frac{1}{2}}$, and $\spin$ is implicit \cite{Cataldi2024Simulating2+1DSU2, Calajo2024DigitalQuantumSimulation, Calajo2025QuantumManybodyScarring-2, Cataldi2025DisorderFreeLocalizationFragmentation-1}.
Matter fields have a corresponding staggered mass-energy term with coupling $\mass$. 
The last two terms of \cref{eq_H} represent the (chromo) electric and magnetic gauge-field energies coupled with $\gauge[E]\!=\!\frac{\gauge[0]^{2}}{2\lspace}$ and $\gauge[B]\!=\!\frac{1}{2\gauge[0]^{2}\lspace}$, respectively, where $\gauge[0]$ is the gauge coupling.

Here, we express gauge link states in the chromoelectric basis $\ket{\spin, \mL, \mR}$, where $\spin\!\in\!\mathbb{N}/2$ indicates the spin irreps and $\mR,\mL\in\{-\spin,\ldots,+\spin\}$ label the states within the spin shell $\spin$.
On this basis, the link energy density operator is diagonal and coincides with the quadratic Casimir, $\hat{E}^2\ket{\spin, \mL, \mR}\!=\!\spin(\spin\mathop+1)\ket{\spin, \mL, \mR}$ \cite{Zohar2015FormulationLatticeGauge}.
Correspondingly, the magnetic energy density is expressed in terms of a four-body plaquette operator $\plaq\!=\! \sum_{\alpha,\beta,\gamma,\delta}\NApara_{\vecsite, \latvec[x]}\Apara_{\vecsite+\latvec[x],\latvec[y]}^{\beta\gamma}\Apara_{\vecsite+\latvec[y],\latvec[x]}^{\gamma\delta\dagger}\Apara_{\vecsite,\latvec[y]}^{\delta\alpha\dagger}$. In this study, we truncate the infinite-dimensional gauge fields \cite{Cataldi2024Simulating2+1DSU2} to $\spin\!\in\!\qty{0,\jonehalf, 1}$ retaining the states that can be reached from the singlet $\ket{\spin\!=\!0,\mL\!=\!\mR\!=\!0}$ by applying the parallel transporter $\hat{U}^{\alpha\beta}$ at most twice.
Such an approximation is perfectly reasonable in the confined phase, where strings and their breaking are well-defined. 

Non-Abelian $\mathrm{SU}(2)$ gauge invariance is then locally imposed with SU$(2)$ generators $\hat{\mathcal{G}}_{\site}\!=\!(\hat{\mathcal{G}}^{x}_{\vecsite},\hat{\mathcal{G}}^{y}_{\vecsite},\hat{\mathcal{G}}^{z}_{\vecsite})$ satisfying $\hat{\mathcal{G}}_{\vecsite}\ket{\Psi_{\qty{b_{\vecsite}}}}=b_{\vecsite}\ket{\Psi_{\qty{\vb{b}_{\site}}}},\,\forall \vecsite \in \Lambda$,
where $b_{\vecsite}$ is the background charge at lattice site $\vecsite$. Using a dressed-site approach \cite{Cataldi2024Simulating2+1DSU2, Calajo2024DigitalQuantumSimulation, Calajo2025QuantumManybodyScarring-2, Cataldi2025DisorderFreeLocalizationFragmentation-1}, we defermionize \cite{Ballarin2024DigitalQuantumSimulation} the original LGT and build a local Hilbert space that retains only those states that satisfy Gauss's law. 
In the \emph{hardcore-gluon} approximation (\idest{}, $\jmax=\spin_{b}=\frac{1}{2}$), we have a $30$-dimensional local basis for each site with zero background charge $\ket{\spin_{b}\!=\!0,m_{b}\!=\!0}$, and with a $42$-dimensional local basis on each of the two lattice sites where we place the static background charges $\qty{\ket{\spin_{b}\!=\!\frac{1}{2},m_{b}\!=\!\pm \frac{1}{2}}}$. 
When $\jmax=\spin_{b}=1$, we have a $317$-dimensional basis for each site with a static background charge and a $168$-dimensional basis for each site with zero background charge; see Supplemental Material (SM) \cite{SM}. 

\begin{figure}
\includegraphics[width=1\columnwidth]{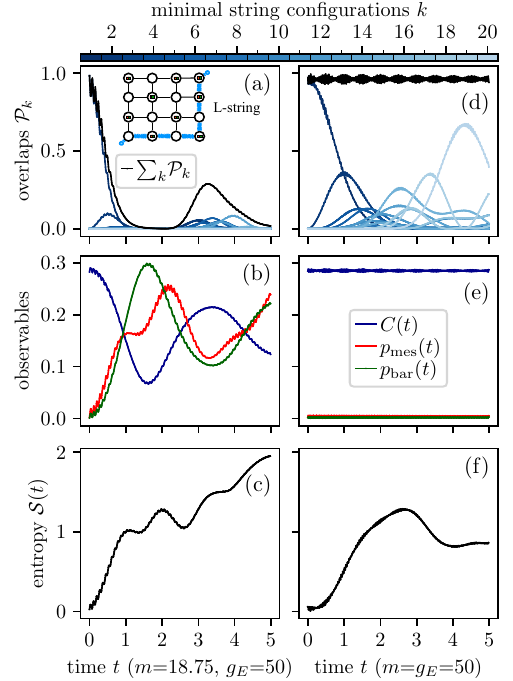}
\caption{\textbf{String breaking vs.~oscillations.} String dynamics starting from a minimal L-string.
\emph{String breaking} at the resonance $\mass\!=\!\tfrac38\,\gauge[E]$ manifest in the (a) decay of string-manifold overlaps over time, (b) decay of the Casimir and increase in the populations of mesons and baryons, defined in \cref{eq_local_observables}, and (c) a rapid continual growth in the entanglement entropy. 
Out of resonance, persistent \emph{string oscillations} appear where (d) the probability of being in a string is always close to unity, (e) matter creation is suppressed, and (f) entanglement entropy saturates at later times.
}
\label{fig2_stringbreaking}
\end{figure}

\textbf{\textit{String initialization.---}}To study the real-time string dynamics of the model, we focus on the strong-coupling regime of \cref{eq_H}, $\gauge[E],\mass\!\gg\!\gauge[B],w$, where the electric term dominates and spontaneous matter creation and plaquette dynamics are suppressed.
On top of a \emph{bare vacuum} configuration with empty even and doubly occupied odd sites \cite{Calajo2025QuantumManybodyScarring-2}, we set the initial string state $\ket{\psi_{0}}$ placing two static SU$(2)$ background charges $\spin_b\!=\!\jonehalf$ within a patch centered in the lattice bulk and connecting them by a path of active $\spin\!=\!\jonehalf$ links (see \cref{fig1_scheme} for a sketch).

Then, to detect string breaking, we measure the overlap $\mathcal{P}_{k}{=}\lvert\braket{\psi_{k}|\psi(t)}\rvert^{2}$ of the time-evolved wave function $\ket{\psi(t)}{=}e^{-i\hat{H}t}\ket{\psi_0}$ with any flux configuration $\ket{\psi_k}$ of the same length as the string of the initial state $\ket{\psi_0}$. Correspondingly, we track the expectation value of local observables $\avg{\hat{O}}\!=\!\expval{\hat{O}}{\psi(t)}$ at any time $t$, properly normalized by the number of lattice sites ($\Nsites[\rm{patch}]$) or links ($\mathcal{L}_{\rm{patch}}$) in the patch of the string.
In particular, we focus on the quadratic Casimir operator and particle densities, such as mesons and (anti)baryons:
\begin{subequations}
\label{eq_local_observables}
\begin{align}
C&\!=\!\frac{1}{\mathcal{L}_{\rm{patch}}}\sum_{\vecsite, k}\avg{\hat{E}^2_{\vecsite,+\latvec[k]}}, \;\; p_{\rm{mes}}\!=\!\frac{1}{\Nsites[\rm{patch}]}\sum_{\vecsite}\avg{\hat{\rho}_{\vecsite}^{[1]}},\\
p&_{\rm{bar}}\!=\!\frac{1}{2\Nsites[\rm{patch}]}\!\sum_{\vecsite}\!\qty[
1\!+\!(\!-\!1)^{\site[x]+\site[y]}
\qty(\avg{\hat{\rho}_{\vecsite}^{[2]}}\!-\!\avg{\hat{\rho}_{\vecsite}^{[0]}})],
\end{align}
\end{subequations}
which are defined by means of the local number occupation operators: (zero) $\avg{\hat{p}_{\vecsite}^{[0]}}\!=\!1\!-\!\avg{\hat{p}_{\vecsite}^{[1]}}\!-\!\avg{\hat{p}_{\vecsite}^{[2]}}$, (single)  $\avg{\hat{p}_{\vecsite}^{[1]}}\!=\!\sum_{\alpha}\avg{\hpsi_{\vecsite,\alpha}^{\dagger}\hpsi_{\vecsite,\alpha}}\!-\!2\avg{\hat{p}_{\vecsite}^{[2]}}$, and (double) $\avg{\hat{p}_{\vecsite}^{[2]}}\!=\!\avg{\hpsi_{\vecsite,\rla}^{\dagger}\hpsi_{\vecsite,\rla}\hpsi_{\vecsite,\gla}^{\dagger}\hpsi_{\vecsite,\gla}}$.
Ultimately, we measure the bipartite entanglement entropy $\entropy\!=\!\!-\!{\Tr}[\hat{\density}_A \log \hat{\density}_A]$, where $\hat{\density}_{A}$ is the reduced density operator of half the system.
Throughout this work, all the results are obtained from TTN simulations of time evolution under \cref{eq_H} on an $8\!\times\!8$ lattice and a $4\!\times\!4$ patch setting the couplings $\gauge[B]=w=1$. 

\textbf{\textit{String breaking.---}}In the strong-coupling regime, the string energy is well approximated by the number of active links $\mathcal{L}_{\rm{active}}$ times their electric cost:
\begin{equation}
    E(t=0)=\langle\psi_{0}|\ham| \psi_{0}\rangle\sim \spin(\spin+1)\gauge[E]\mathcal{L}_{\rm{active}}.
    \label{eq_en_initial_state}
\end{equation}
Under unitary time evolution, (first-order) string breaking for $\spin\!=\!\frac{1}{2}$ is enabled when the energy $\spin(\spin+1)\gauge[E]\!=\!\frac{3}{4}\gauge[E]$ of removing one active link matches the energy $2\mass$ of creating a particle-hole pair at the sites connecting it.
At this resonance point $R_{\spin=\frac{1}{2}}{:}\, \mass\!=\!\frac{3}{8}\gauge[E]$, any string configuration of length equal to that of the initial string resonantly hybridizes with shorter strings carrying dynamical matter: the Casimir along the path decreases, the total \emph{string-manifold} weight $\mathcal{P}=\sum_{k}\mathcal{P}_{k}$ (sum over degenerate minimal-string configurations) collapses, and matter production proliferates, leading to a rapid and continual growth in the bipartite entanglement entropy [see \cref{fig2_stringbreaking}(a,b,c) for a minimal initial L-string].

Unlike the Abelian case \cite{Xu2025StringBreakingDynamics}, matter production at resonance is not limited to mesons: SU$(2)$ fusion rules also enable spontaneous \emph{baryon-antibaryon} formation, visible in \cref{fig2_stringbreaking}(c). 
Moreover, due to the use of staggered fermions, higher-order string breaking processes (\idest{}, multiple local breaks) can occur at the same resonance $R_{\spin=\frac{1}{2}}$. Videos of this dynamics are available on \texttt{YouTube} \cite{videos}.

\begin{figure}
\includegraphics[width=1\columnwidth]{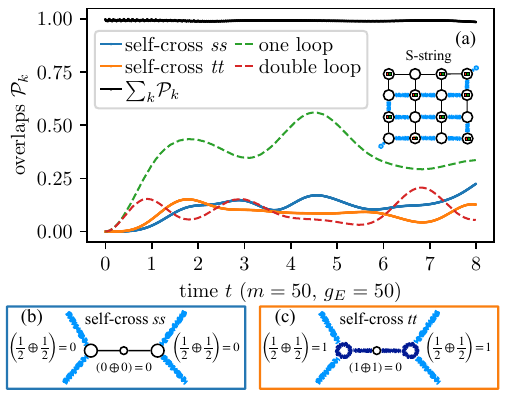}
\caption{\textbf{Glueball and self-crossing string formation.} 
String oscillations starting from a maximal S-string.
Among all the degenerate string configurations $\mathcal{P}_{k}$ (whose sum stays close to unity) available out of resonance, we highlight in (a) the superpositions of type-\emph{ss} and type-\emph{tt} self-crossing strings and loop configurations (glueballs) of single and double loops, respectively. 
The lower panel shows how four $\spin=\jonehalf$ (cyan) gauge links can fuse into a singlet using the SU$(2)$ intertwiner scheme: the \emph{ss}-type (b), where each link pair intermediately fuses to $\spin=0$ (black), and the \emph{tt}-type where each link pair intermediately fuses to $\spin=1$ (dark blue).}
\label{fig3_glueballs}
\end{figure}

\textbf{\textit{String oscillations: glueballs and self-cross.---}}Out of resonance, \idest{}, for $\mass\!\neq\!\tfrac38\gauge[E]$, the initial string cannot break; the plaquette term only reshuffles flux among configurations with the same number of active links.
Accordingly, the patch-averaged Casimir remains constant without particle creation. 
At the same time, the entanglement entropy peaks near maximal superposition of the degenerate string configurations and minimizes when a single configuration dominates; see \cref{fig2_stringbreaking}(d,e,f). 

For initial strings of nonminimal length, like the \emph{snake} S-string in \cref{fig3_glueballs}, the magnetic term generates fluctuations into shorter strings, pure-gauge loops (\emph{glueballs}) \cite{Mathieu2009PhysicsGlueballs}, and \emph{self-crossed} strings. 
While glueballs and self-crossings also appear in Abelian $\mathbb{Z}_2$ LGTs \cite{Xu2025StringBreakingDynamics}, the SU$(2)$ self-crossing case is richer because a four-valent vertex with four $\spin=\frac{1}{2}$ links supports two nonequivalent singlet intertwiners \cite{Zache2023QuantumClassicalSpinNetwork}.
We label them: type-\emph{ss}, where the two link pairs fuse as singlets $(\tfrac12\!\otimes\!\tfrac12)\!\to\!0$ and then $(0\!\otimes\!0)\!\to\!0$ [see \cref{fig3_glueballs}(b)]; and type-\emph{tt}, where the pairs first fuse as triplets $(\tfrac12\!\otimes\!\tfrac12)\!\to\!1$ and then $(1\!\otimes\!1)\!\to\!0$ [see \cref{fig3_glueballs}(c)]. 
Despite being completely equivalent in terms of Hamiltonian interactions, the two intertwiners display different dynamics, with the type-\emph{ss} slightly dominating in time; see \cref{fig3_glueballs}(a). 
Such a discrepancy might originate from plaquette flips acting on one link pair of such configurations: the two $\spin\!=\!\tfrac12$ links preferentially reduce to $\spin\!=\!0$, naturally feeding the \emph{ss}-channel and depriving type-\emph{tt} of resonant pathways.  Videos of this dynamics are available on \texttt{YouTube} \cite{videos}.

\begin{figure}[!t]
\includegraphics[width=1\columnwidth]{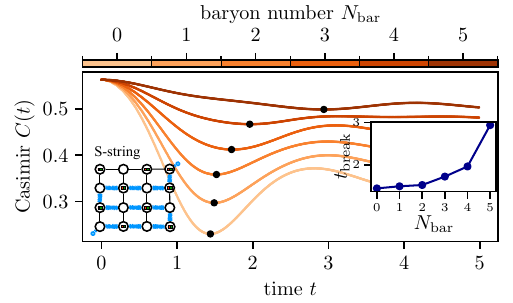}
\caption{
\textbf{Baryon blockade.} 
Effect of the baryon number sector $N_{\rm{bar}}$ on the string-breaking mechanism in terms of the first observed minimum ($t_{\rm{break}}$) in the time evolution of the Casimir operator. 
The inset panel shows the scaling of $t_{\rm{break}}$ as a function of the baryon number sector $N_{\rm{bar}}$.}
\label{fig4_baryon_blockade}
\end{figure}

\begin{figure}[!t]
\includegraphics[width=1\columnwidth]{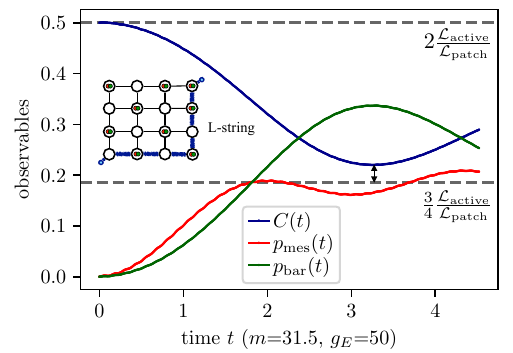}
\caption{
\textbf{String tension reduction mechanism.} 
At the resonance $R_{j=1}: \mass\!=\!\tfrac58\gauge[E]$, the $\spin\!=\!1$ initial L-string reduces its tension from $C=2\tfrac{\mathcal{L}_{\rm{active}}}{\mathcal{L}_{\rm{patch}}}$ (upper dashed line) towards $C\gtrsim\tfrac{3}{4}\tfrac{\mathcal{L}_{\rm{active}}}{\mathcal{L}_{\rm{patch}}}$ (lower dashed line) through meson and baryon creation while not being broken.}
\label{fig5_tension_reduction}
\end{figure}

\textbf{\textit{Baryon blockade mechanism.---}}Unlike MC simulations, where finite (baryon) density regimes are challenging because of the sign problem \cite{Nagata2022FinitedensityLatticeQCD}, here the baryon number is a quantum number associated with the global particle-hole U$(1)$ symmetry of \cref{eq_H}, and can be easily fixed in exact diagonalization (ED) \cite{Cataldi2024EdlgtExactDiagonalizationLattice} and TN simulations \cite{Silvi2014LatticeGaugeTensor, Silvi2019TensorNetworksAnthology}.
It is then interesting to investigate the role of finite baryon density on the dynamical string breaking mechanism.

A good measure for the string breaking is the break time $t_{\rm break}$ corresponding to the minimum of the average Casimir operator. 
Indeed, as previously shown in \cref{fig2_stringbreaking}(a,b,c), the Casimir minimum occurs in parallel to the minimum of the string-manifold weight $\mathcal{P}$ and the peak of meson and baryon formation (see also \cite{SM}).

As shown in \cref{fig4_baryon_blockade} for an initial S-string, adding SU$(2)$ baryons at the resonance $\mass=\tfrac38\gauge[E]$ progressively delays $t_{\rm{break}}$.
We attribute this to a \emph{baryon blockade} effect: finite density reduces the availability of final states for meson creation and prunes plaquette-mediated resonant paths, suppressing the string breaking mechanism.

\textbf{\textit{String tension reduction.---}}The \emph{hardcore-gluon} truncation at $\jmax\!=\!\jonehalf$ and the minimal background charge $j_b\!=\!\jonehalf$ in the confinement phase already captures pure non-Abelian string dynamics. 
Enabling larger gauge-field truncations (and static charges) enriches the physics by opening \emph{representation-changing} resonances where the string \emph{reduces} its tension without fully breaking.
Consider an initial string with $\spin\!=\!1$, i.e., $\spin(\spin\!+\!1)\!=\!2$, between two static charges with $\spin_{b}\!=\!1$, whose energy still follows \cref{eq_en_initial_state}.
Then, a single particle-hole pair creation on neighboring sites along the path can reduce a link from $\spin\!=\!1$ to $\spin\!=\!\tfrac12$, lowering its chromo-electric energy by $\Delta E(1\to \tfrac12)\!=\!\qty(2-\tfrac34)\gauge[E]\!=\!\tfrac54\gauge[E]$ at the cost of $2\mass$. 
Crucially, such a process only occurs at the \emph{tension-reduction resonance} $R_{\spin=1}\!:\mass\!=\!\tfrac58\gauge[E]$ and is not compatible with the subsequent breaking ($\tfrac12\!\to\!0$), whose energy cost \(\Delta E(\tfrac12\!\to\!0)\!=\!\tfrac34\,\gauge[E]\) does not match $2\mass\!=\!\tfrac54\gauge[E]$. 
Hence, the string does not fully break; it relaxes at lower tension ($\spin\!=\!\jonehalf$), as further shortening requires off-resonant (plaquette-assisted) or higher-order processes. 

As shown in \cref{fig5_tension_reduction} for an initial L-string at $R_{\spin=1}$, tension reduction manifests as a reduction of the patch-averaged Casimir ($2\,\gauge[E]\to\tfrac34\,\gauge[E]$) accompanied by meson and baryon-antibaryon production. In general, for arbitrarily large-$\spin$ strings, (first-order) string-tension reduction $(\spin\to\spin\!-\!\jonehalf)$ occurs for a family of resonances $R_j\!: 2\mass=\gauge[E]\qty(\spin+\tfrac14)$, which coincides with string breaking only when $\spin\!=\!\tfrac12$.

\textbf{\textit{Summary and outlook.---}} We have performed the first TN real-time simulations of string dynamics in a $2{+}1$D SU$(2)$ Yang--Mills LGT with dynamical matter.
In a strong-coupling regime, we have identified a resonance where string states anticross with broken-string manifolds, forming mesons and baryons.  
We uncovered a many-body, non-Abelian effect—baryon-blockade string breaking—where finite baryon density delays string breaking at fixed resonance. 
Out of resonance, the magnetic term generates pure non-Abelian glueballs and self-crossed strings that resolve two SU$(2)$ intertwiners at a four-valent vertex, with type-\emph{ss} favored by electric-energy detuning and plaquette flips. 
Finally, at higher gauge field truncations, we observe representation-dependent \emph{tension reduction} $\qty(j\!\to\!j-\tfrac12)$ resonances.

Together, these results provide \textit{ab initio}, real-time, many-body signatures of confinement dynamics in two spatial dimensions, beyond the reach of MC methods. It is important to note that the evolution times and system sizes we access in our TN simulations go significantly beyond the state-of-the-art quantum simulation experiments on string dynamics in $2+1$D LGTs \cite{Cochran2025VisualizingDynamicsCharges, Gonzalez-Cuadra2025ObservationStringBreaking, Cobos2025RealTimeDynamics2+1D}. 
Furthermore, several features of the phenomenology uncovered in this work are distinctively non-Abelian, which not only goes beyond the accessible physics of the aforementioned experiments, but also beyond previous TN works since we probe far-from-equilibrium non-Abelian string dynamics in $2+1$D. 
Consequently, our results push the state of the art in TN simulations of higher-dimensional non-Abelian LGTs and delineate concrete targets for near-term quantum simulators of these phenomenologically rich models.

\medskip
\footnotesize 
\textbf{\textit{Acknowledgments.---}}The authors acknowledge fruitful discussions with Christian Bauer, Umberto Borla, Anthony Ciavarella, Lukas Ebner, Klaus Liegener, Jesse J. Osborne, N. S. Srivatsa, Yizhuo Tian, and Kaidi Xu.
The authors acknowledge funding by the Max Planck Society, the Deutsche Forschungsgemeinschaft (DFG, German Research Foundation) under Germany’s Excellence Strategy – EXC-2111 – 390814868, and the European Research Council (ERC) under the European Union’s Horizon Europe research and innovation program (Grant Agreement No.~101165667)—ERC Starting Grant QuSiGauge. Views and opinions expressed are, however, those of the author(s) only and do not necessarily reflect those of the European Union or the European Research Council Executive Agency. Neither the European Union nor the granting authority can be held responsible for them. This work is part of the Quantum Computing for High-Energy Physics (QC4HEP) working group.  
All the ED and TN numerical simulations have been obtained with \textrm{ed-lgt} \cite{Cataldi2024EdlgtExactDiagonalizationLattice} and Quantum Tea \cite{Baccari2025QuantumTEAQtealeaves, Baccari2025QuantumTEAQredtea}, respectively.
\normalsize

\bibliography{references}

\clearpage
\onecolumngrid
\begin{center}
    \textbf{\large Supplemental Online Material for \\``Real-Time String Dynamics in a \texorpdfstring{$2+1$}{2+1}D Non-Abelian Lattice Gauge Theory:\\ String Breaking, Glueball Formation, Baryon Blockade, and Tension Reduction''}\\[5pt]
    \vspace{0.1cm}
    \begin{quote}
    {\small In this Supplemental Material, we detail the model derivation and the numerical techniques employed in this study.}\\[10pt]
    \end{quote}
\end{center}
\setcounter{equation}{0}
\setcounter{figure}{0}
\setcounter{table}{0}
\setcounter{page}{1}
\setcounter{section}{1}
\makeatletter
\renewcommand{\theequation}{S\arabic{equation}}
\renewcommand{\thefigure}{S\arabic{figure}}
\renewcommand{\thesection}{S\Roman{section}}
\renewcommand{\thepage}{\arabic{page}}
\renewcommand{\thetable}{S\arabic{table}}
\vspace{0cm}
\twocolumngrid
\normalsize
\section{Dressed-site formulation of the SU\texorpdfstring{$(2)$}{(2)} LGT with background charges}
To map the Kogut--Susskind $\mathrm{SU}(2)$ lattice Yang--Mills Hamiltonian of \cref{eq_H} to the dressed-site formulation adopted in numerical simulations, we follow the prescription developed in \cite{Cataldi2024Simulating2+1DSU2} and lately replicated in \cite{Calajo2024DigitalQuantumSimulation, Calajo2025QuantumManybodyScarring-2, Cataldi2025DisorderFreeLocalizationFragmentation-1}. 
In detail, we:
\begin{enumerate*}
    \item decompose the $\mathrm{SU}(2)$ gauge fields in terms of a direct sum of irreps $\spin $;
    \item truncate the links representation up to a fixed $\spin_{\max}$ \cite{Cataldi2024Simulating2+1DSU2};
    \item build the gauge-singlet local dressed basis accounting for the presence of eventual background charges; and 
    \item gauge-defermionize the theory.
\end{enumerate*}

\subsection{Decomposition of matter and gauge fields}
We start by decomposing both matter and gauge local Hilbert spaces in $\mathrm{SU}(2)$ irreps. 
We adopt the irrep basis $\ket{\spin, \mL, \mR}$ for gauge links, and the Fock basis
\begin{math}
    \{\ket{q}\}
    =
    \{
    {\ket{0}},\,
    {\ket{\rla},\ket{\gla} = \hpsi^{\dagger}_{\rla,\gla}\ket{0}},\,
    {\ket{2}    = \hpsi^{\dagger}_{\rla} \hpsi^{\dagger}_{\gla}\ket{0}}
    \}
\end{math}
for matter sites.
Matter basis states are associated with the following spin labels $(j,m)$:
\begin{equation}
    \ket{0}\leftrightarrow (0,0)
    \,\quad
    \ket{\rla},\ket{\gla}\leftrightarrow (0,\pm1/2)
    \,\quad
    \ket{2}\leftrightarrow (0,0)
    \,.
\end{equation}
Irrep decomposition specifies how local gauge rotations act on a site $\vecsite$ and its neighboring links.
The generators of the infinitesimal rotations read $\forall \nu\in \{x,y,z\}$:
\begin{equation}
    \hat{\mathcal{G}}^{\nu}_{\vecsite}=\hat{R}_{\vecsite-\latvec,\vecsite}^{\nu} + \hat{Q}_{\vecsite}^{\nu}+\hat{L}_{\vecsite,\vecsite+\latvec}^{\nu}\,,
\end{equation}
where
\begin{math}
    \hat{Q}^{\nu}_{\vecsite}\!=\!
    \sum_{\alpha\beta}
    \hpsi^{\dagger}_{\vecsite,\alpha}
    S^{(1/2)\nu}_{\alpha\beta}
    \hpsi_{\vecsite,\beta}
\end{math}
rotates the quark field at $\vecsite$, while $\hat{R}_{\vecsite-\latvec,\vecsite}^{\nu}$ and $\hat{L}^{\nu}_{\vecsite,\vecsite+\latvec}$ account for the transformation of the gauge links at its left and right \cite{Zohar2015FormulationLatticeGauge}:
\begin{align}
\label{eq_LR_rishon_rotations}
        \mel{\spin^{\prime} \mL^{\prime} \mR^{\prime}}{\hat{L}^{\nu}}{ j \mL \mR } & =\delta_{\spin,\spin^{\prime}}S^{(j)\nu}_{\mL^{\prime},\mL} \delta_{\mR^{\prime},\mR}\,,\\
        \mel{\spin^{\prime} \mL^{\prime} \mR^{\prime} }{\hat{R}^{\nu}}{j \mL \mR } & =\delta_{\spin,\spin^{\prime}}\delta_{\mL^{\prime},\mL} S^{(j)\nu}_{\mR^{\prime},\mR}\,;
\end{align}
where $S^{(j)\nu}$ are the spin-$\spin$ $\mathfrak{su}(2)$ matrices.
From these operators, we build the (chromo)electric energy operator 
\begin{math}
    \hat{E}^2 =\sum_{\nu}(\hat{R}^{\nu})^2 =\sum_{\nu}(\hat{L}^{\nu})^2
\end{math},
\idest{}, quadratic Casimir \cite{Zohar2015FormulationLatticeGauge}:
\begin{align}
    \hat{E}^2\ket{\spin \mL \mR }&=\spin(\spin+1)\ket{\spin \mL \mR }\,.
\end{align}
Correspondingly, the action of the parallel transporter $\hat{U}_{\alpha,\beta}$ is given in terms of Clebsch-Gordan coefficients \cite{Zohar2015FormulationLatticeGauge}:
\begin{equation}
\label{eq_parallel_transporter}
    \mel{\spin^{\prime}\mL^{\prime}\mR^{\prime}}{\hat{U}^{\alpha\beta}}{\spin\mL \mR} 
    =\sqrt{\frac{2\spin+1}{2\spin ^{\mathrlap{\prime}}+1}}\:
    \overline{C^{\spin,\mL}_{\jonehalf,\alpha;\,\spin^{\prime},\mL^{\prime}}}
    C^{\spin^{\prime},\mR^{\prime}}_{\jonehalf,\beta;\,\spin,\mR}.
\end{equation}
Then, the local $\mathrm{SU}(2)$ gauge-invariance of \cref{eq_H} reads
\begin{equation}
    \hat{\mathcal{G}}_{\vecsite}^{\nu}
    \ket{\Psi_{\qty{b_{\vecsite}}}} = 
    b_{\vecsite}^{\nu}
    \ket{\Psi_{\qty{\vb{b}_{\vecsite}}}}, \; \forall \vecsite
\end{equation}
where $b_{\vecsite}^{\nu}$ is the static background-charge present on the lattice site $\vecsite$ in the QMB state $\ket{\Psi}$.

\subsection{Gauge field and background charge truncations}
Within the irrep basis, gauge fields and static background charges can occupy arbitrarily high spin shells.
To deal with a finite gauge-link Hilbert space, we consider a maximal irrep $\jmax$ which yields an energy cutoff $\norm*{\hat{E}^2} \leq (\jmax(\jmax +1))$ on the Casimir spectrum ($\norm{\:\cdot\:}$ denotes the matrix norm) \cite{Cataldi2024Simulating2+1DSU2, Calajo2024DigitalQuantumSimulation, Calajo2025QuantumManybodyScarring-2, Cataldi2025DisorderFreeLocalizationFragmentation-1}.
A similar truncation is performed on the irreps of the static background charges $(\spin_{b},m^{b})$, where $\spin_{b}\leq \jmax^{b}$.

Recovering the full, untruncated gauge group is \emph{not} essential for the present investigation. 
Indeed, all the discussed string dynamics phenomena have been studied in the strong coupling regime $\gauge[E], m \gg \gauge[B],t$, where the chromoelectric energy and the mass term dominate the Hamiltonian, suppressing any gauge irrep fluctuations arising from the plaquette term or the hopping interaction. 
In this regime, even the simplest nontrivial approximation $\jmax\!=\!\jmax^{b}=1/2$, called \emph{hardcore-gluon} approximation, faithfully captures all the predicted string phenomena in and out of resonance.
As shown in the main text, considering larger irrep truncations $\spin{\to} \spin^{\prime}\!=\!\spin\!+\!1/2$ generalizes the string breaking within string-tension reduction occurring when $\gauge[E](\spin^{\prime}(\spin^{\prime}+1) = \gauge[E](\spin(\spin+1)+2m$. 

\subsection{Dressed-site basis with background charges}
Regardless of the gauge-field and the background-charge truncations, we observe that $\hat{L}^{\nu}$ and $\hat{R}^{\nu}$ from \cref{eq_LR_rishon_rotations} act nontrivially only on the $\mL$ and $\mR$ index, respectively.
We can then factorize each gauge link in a pair of rishon degrees of freedom, living at its edges \cite{Silvi2014LatticeGaugeTensor}.
Combining each matter site with its two adjacent (left and right) gauge-rishons and the attached background charge, we forge a composite $d$-dimensional dressed-site basis where Gauss's law is automatically satisfied \cite{Cataldi2024Simulating2+1DSU2}.
Moreover, since $\forall \vecsite$, $[\hat{\mathcal{G}}_{\vecsite}\!-\!b_{\vecsite},\ham]=0$, each background-charge sector is completely decoupled and does not interact with each other \cite{Cataldi2025DisorderFreeLocalizationFragmentation-1}.
This guarantees that any background charge placed in the lattice remains static and unaffected by the system dynamics.

Within the \emph{hardcore-gluon} truncation \cite{Cataldi2024Simulating2+1DSU2, Calajo2024DigitalQuantumSimulation, Calajo2025QuantumManybodyScarring-2, Cataldi2025DisorderFreeLocalizationFragmentation-1}, corresponding to $\jmax\!=\!\jmax^{b}\!=\!1/2$, 
we have a $d\!=\!72$-dimensional local dressed-site Hilbert space, which is divided into two sectors identified by the value of the background charge: the first 30 states of the basis belong to the gauge-invariant sector with zero background-charge $b_{0}\!=\!\ket{\spin_{b}=0,m_{b}=0}$ \cite{Cataldi2024Simulating2+1DSU2}, while the remaining 42 states belong to the sector with a background-charge $b_{\jonehalf}\!=\!\ket{\spin_{b}=1/2,m_{b}\in\qty{+1/2, -1/2}}$ \cite{Cataldi2021HilbertCurveVs} and corresponds to the sites hosting the edges of the string configurations investigated in this study.
Similarly, when extending the truncation up to $\jmax\!=\!1\!=\!\jmax^{b}$, all the lattice sites with no background charge have a dressed-site Hilbert space with $168$ states (see Tab.~1 of \cite{Magnifico2025TensorNetworksLattice}), while the two sites hosting the edges of the string have a Hilbert space with $317$ states with a background charge $b_{1}\!=\!\ket{\spin_{b}=1,m_{b}\in\qty{+1,0,-1}}$.

\subsection{Rishon decomposition of gauge fields}
The dressed-site formalism previously introduced is equivalent to the original LGT description in \cref{eq_H} as long as it recovers the original physical space. 
This requires that the left and right rishons on each link, identifying the left and right gauge link components, must be in the same spin irrep $\spin$ (\idest{} must have the same Casimir). 
Namely, $\forall \vecsite, \latvec \in \Lambda$, it must hold:
\begin{equation}
\label{eq_link_symmetry}
\sum_{\nu}\qty(\hat{L}_{\vecsite,\latvec}^{\nu})^{2} = \sum_{\nu}\qty(\hat{R}_{\vecsite+\latvec,-\latvec}^{\nu})^{2}.
\end{equation}
This condition can be easily imposed as a two-body U$(1)$ symmetry via ED \cite{Cataldi2024EdlgtExactDiagonalizationLattice} or TN \cite{Baccari2025QuantumTEAQtealeaves} simulation and is preserved during the system's time evolution.
Correspondingly, the parallel transporter in \cref{eq_parallel_transporter} can be decomposed into its action on the left and the right halves of the gauge link, named rishons, to which we can associate a fermionic statistics (see \cite{Ballarin2024DigitalQuantumSimulation, Cataldi2024Simulating2+1DSU2, Cataldi2025HamiltonianLatticeGauge}). 

The general Rishon decomposition of $\hat{U}^{\alpha\beta}$ for an arbitrary truncation of the maximum allowed spin shell $\jmax$ is a bilinear form, which separately accounts for the action of both rishons increasing the irrep $\spin\to \spin+\jonehalf$, and both decreasing the irrep $\spin\to \spin-\jonehalf$. Namely \cite{Cataldi2024Simulating2+1DSU2}:
\begin{equation}
    \hat{U}^{\alpha\beta}_{\vecsite, \vecsite+\latvec}=\rishon_{A,\vecsite,\latvec}^{\alpha} \rishon_{B,\vecsite+\latvec,-\latvec}^{\beta\dagger}+\rishon_{B,\vecsite,\latvec}^{\alpha} \rishon_{A,\vecsite+\latvec,-\latvec}^{\beta\dagger},
    \label{eq_SU2_U_definition}
\end{equation}
where the two $\rishon$-rishon species, A and B, act respectively as raising and lowering the spin shell of the SU$(2)$ gauge irrep.
Moreover, they are related to each other as:
\begin{align}
    \rishon_{A}^{\alpha} & = i \sigma^{y}_{\alpha,\beta} \rishon_{B}^{\beta\dagger}, &
    \rishon_{A}^{\alpha\dagger} = i \sigma^{y}_{\alpha,\beta} \rishon_{B}^{\beta}.
\end{align}
We can then rewrite \cref{eq_SU2_U_definition} just in terms of one species, \eg, B.
Dropping the index, i.e. $\rishon_{B}^{\alpha}=\rishon^{\alpha}$, it holds:
\begin{equation}
    \hat{U}^{\alpha\beta}_{\vecsite, \vecsite+\latvec}=
    i \sigma^{y}_{\alpha,\gamma} \rishon_{\vecsite,\latvec}^{\gamma\dagger} \rishon_{\vecsite+\latvec,-\latvec}^{\beta\dagger}
    + i \sigma^{y}_{\beta,\gamma} \rishon_{\vecsite,\latvec}^{\alpha} \rishon_{\vecsite+\latvec,-\latvec}^{\gamma}.
\end{equation}
For a chosen truncation $\jmax$ of the SU$(2)$ irrep, $\zeta$-rishons are defined as follows \cite{Cataldi2024Simulating2+1DSU2}:
\begin{equation}
    \rishon^{\gla(\rla)}\!=\!\qty[\sum_{j=0}^{\jmax-\jonehalf}\sum_{m=-j}^{j}\chi(j,m,\gla(\rla)) \ket{j,m}\bra{j+\mbox{$\jonehalf$},m{\cmp}\mbox{$\jonehalf$}}],
    \label{eq_SU2_general_rishon}
\end{equation}
where the function $\chi(j,m,\alpha)$ reads
\begin{equation}
    \chi\qty(j,m,\gla(\rla))=\sqrt{\frac{j\cmp m+1}{\sqrt{(2j+1)(2j+2)}}}.
\end{equation}
In the \emph{hardcore-gluon} approximation, the parallel transporter reduces to \cite{Zohar2015FormulationLatticeGauge}: 
\begin{equation}
    \NApara\!=\!{\frac{1}{\sqrt{2}}}
    \qty(
      \begin{array}{@{\hspace{0.1ex}}c@{\hspace{0.1ex}}|@{\hspace{0.1ex}}c@{\hspace{0.2ex}}c@{\hspace{0.2ex}}c@{\hspace{0.2ex}}c@{\hspace{0.1ex}}}
    0&\!+\!\delta_{\alpha\rla}\delta_{\beta\gla} 
     &\!-\!\delta_{\alpha\rla}\delta_{\beta\rla}  
     &\!+\!\delta_{\alpha\gla}\delta_{\beta\gla}
     &\!-\!\delta_{\alpha\gla}\delta_{\beta\rla} \\
    \hline
    \!-\!\delta_{\alpha\gla} \delta_{\beta\rla}&0&0&0&0\\
    \!-\!\delta_{\alpha\gla} \delta_{\beta\gla}&0&0&0&0\\
    \!+\!\delta_{\alpha\rla} \delta_{\beta\rla}&0&0&0&0\\
    \!+\!\delta_{\alpha\rla} \delta_{\beta\gla}&0&0&0&0\\
  \end{array})
\label{SU2_parallel_transport}
\end{equation}
where the $1/\sqrt{2}$ factor ensures that the hopping term preserves the state norm on its support. 
Similarly, also the rishon-decomposition in \cref{eq_SU2_U_definition} simplifies to:
\begin{equation}
    \NApara_{\genlink}=
    \rishon_{\genlink}^{\alpha} \rishon_{\siteplus,-\latvec}^{\beta\dagger}\,,
\end{equation}
where $\rishon$-rishons in \cref{eq_SU2_general_rishon} reduce to:
\begin{equation}
    \begin{aligned}
        \rishon_{\rla} &
        =\frac{1}{\sqrt[4]{2}} \qty( 
        \begin{array}{c|cc}
            0 & 1 & 0 \\
            \hline
            0 & 0 & 0 \\
            1 & 0 & 0 \\  
        \end{array})&
        \rishon_{\gla} &
        =\frac{1}{\sqrt[4]{2}} \qty( 
        \begin{array}{c|cc}
            0 & 0 & 1 \\
            \hline
            \!-\!1 & 0 & 0 \\
            0 & 0 & 0 \\  
        \end{array}),
    \label{zeta_definition}
    \end{aligned}
\end{equation}
It is possible to prove \cite{Cataldi2025HamiltonianLatticeGauge, Cataldi2024Simulating2+1DSU2}
that such a definition of the SU$(2)$-rishons satisfies all the needed properties.
\begin{figure*}[!ht]
\includegraphics[width=1\textwidth]{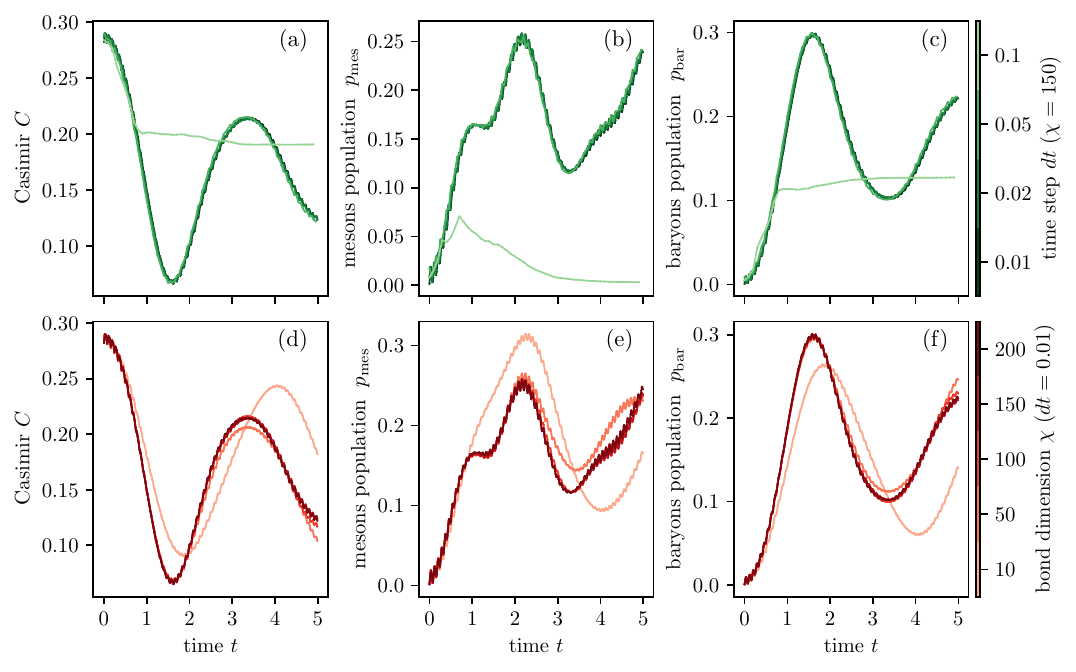}
\caption{\textbf{Convergence in time step and bond dimension} Convergence behavior of the observed string-breaking dynamics as a function of (a)-(c) the time step $dt$ (at fixed bond dimension $\chi\!=\!150$) and (d)-(f) bond dimension $\chi$ (at fixed time step $dt\!=\!0.01$) adopted in the TVDP protocol for TTN time evolution. 
From left to right, plots show the time evolution of (a,d) the Casimir operator, (b,e) the meson population, and (c,f) the baryon population.}
\label{fig_convergence}
\end{figure*}
\subsection{Gauge defermionization}
Within the dressed-site formulation adopted so far, the model still describes degrees of freedom with fermionic (matter fields) and bosonic (gauge fields) statistics. 
While specific quantum hardware \cite{Esslinger2010FermiHubbardPhysicsAtoms, Greif2013ShortRangeQuantumMagnetism, Jordens2008MottInsulatorFermionic, Messer2015ExploringCompetingDensity} and dedicated TN ansätze \cite{Emonts2023FindingGroundState} can handle fermionic statistics, mixing fermions and bosons at a numerical level, especially beyond one dimension, can be very challenging \cite{Ballarin2024DigitalQuantumSimulation}. 
However, the existence of a $\mathbb{Z}_2$-subgroup within the SU($2$) gauge group enables the complete defermionization of the model in \cref{eq_H} \cite{Cataldi2024Simulating2+1DSU2} while dealing with bosonic dressed-site operators.
In practice, we need to associate a fermionic statistic to each $\rishon^{\alpha}$-rishon in \cref{eq_SU2_general_rishon,zeta_definition}, in such a way that they anti-commute among themselves at different orbitals and with matter fields:
\begin{equation}
  \qty{\hat{\zeta}_{\vecsite,\latvec}^{\alpha},\hat{\zeta}_{\vecsite^{\prime},\latvec^{\prime}}^{\beta}}\!=\!0,\;\;
  \qty{\hat{\zeta}_{\vecsite,\vb*{\mu}}^{\alpha},\hat{\psi}_{\vecsite,\beta}}\!=\!0\;\; \forall \alpha,\beta, \vecsite, \latvec.
  \label{eq_SU2_zeta_commmutations}
\end{equation}
For a generic system with fermionic particles arbitrarily sorted along a certain path, a generic fermionic operator $\hat{F}_{\vecsite}$ acting on the $\vecsite^{th}$ position along the path reads
\begin{equation}
    \hat{F}_{\vecsite}=\dots P_{\vb{n-2}}\otimes P_{\vb{n-1}}\otimes F_{\vecsite} \otimes \mathbb{1}_{\vb{n+1}}\otimes \mathbb{1}_{\vb{n+2}}\dots 
    \label{fermionic_qmb_op}
\end{equation}
where $P_{\vecsite}=P_{\vecsite}^{\dagger}=P_{\vecsite}^{-1}$ is a fermion parity operator that gets inverted after the action of a fermionic operator:
\begin{equation}
\qty{P_{\vecsite},F_{\vecsite}}=0 \qquad 
\qty[P_{\vecsite},F_{\vecsite^{\prime}\neq \vecsite}]=0 
\qquad \forall \vecsite,\vecsite^{\prime}\in \Lambda
\label{fermion_parity_commutation}
\end{equation}
Therefore, any matter field admits its own notion of parity satisfying \cref{fermion_parity_commutation}. For Dirac fermions, we have:
\begin{align}
  \hat{\psi}_{\text{Dirac}} &= \qty( 
  \begin{array}{cc}
   0 & 1 \\
   0 & 0
   \end{array})_F&
   \hat{P}_{\text{Dirac}} &= \qty( 
  \begin{array}{cc}
   +1 & 0 \\
   0 & -1
   \end{array})  
\end{align}
where the subscript $F$ reminds that the $\hpsi$ matrix behaves as a fermion, with the global action in \cref{fermionic_qmb_op}.
Similarly, we can define the SU(2) Rishon parity operator $P_{\zeta}$ with an even ($+1$) parity sector on \emph{integer} irreps and odd ($-1$) sector on \emph{semi-integer} ones (see Eq.1.3.30 of \cite{Cataldi2025HamiltonianLatticeGauge}). 
Correspondingly, the parallel transporter $\hat{U}_{\vecsite,\vecsite+\vb*{\mu}}^{\alpha\beta}$ reads:
\begin{equation}
  \begin{split}
    \hat{U}_{\vecsite,\vecsite+\vb*{\mu}}^{\alpha\beta}=& 
 i\sigma^{y}_{\alpha\gamma} \hat{\zeta}_{\vecsite,\vb*{\mu}}^{\gamma\dagger} \hat{\zeta}_{\vecsite+\vb*{\mu},-\vb*{\mu}}^{\beta\dagger} 
 + i\sigma^{y}_{\beta\gamma} \hat{\zeta}_{\vecsite,\vb*{\mu}}^{\alpha} \hat{\zeta}_{\vecsite+\vb*{\mu},-\vb*{\mu}}^{\gamma}\\
    =&+i\sigma^{y}_{\alpha\gamma}\qty(\zeta_{\vecsite,\vb*{\mu}}^{\gamma\dagger}\cdot P_{\zeta, \vecsite,\vb*{\mu}})\otimes \zeta_{\vecsite+\vb*{\mu},-\vb*{\mu}}^{\beta\dagger}\\
    &+i\sigma^{y}_{\beta\gamma}\qty(\zeta_{\vecsite,\vb*{\mu}}^{\alpha}\cdot P_{\zeta, \vecsite,\vb*{\mu}})\otimes \zeta_{\vecsite+\vb*{\mu},-\vb*{\mu}}^{\gamma}
  \end{split}
\end{equation}
In this way, the hopping term of \cref{eq_H} can be rewritten as a two-body interaction of two adjacent dressed site bosonic operators, each one made of a pair of fermions (one matter field and one rishon field).

\section{Tensor Network methods for real-time evolution}

Tensor Network (TN) methods provide efficient representations of quantum many-body states by exploiting the fact that low-energy states of local Hamiltonians typically obey the entanglement area law \cite{Eisert2010ColloquiumAreaLaws, Calabrese2004EntanglementEntropyQuantum}. 
Instead of storing the exponentially large number of coefficients of a generic wavefunction, a TN approximates the state with a polynomial number of parameters, controlled by the bond dimension $\chi$. 
Tuning $\chi$ interpolates between product states ($\chi=1$) and exact representations ($\chi \sim d^N$), making $\chi$ the key parameter for controlling accuracy.

Among the various TN architectures, Tree Tensor Networks (TTNs) \cite{Shi2006ClassicalSimulationQuantum, Silvi2019TensorNetworksAnthology} are particularly suitable for high-dimensional lattice models \cite{Felser2020TwoDimensionalQuantumLinkLattice, Magnifico2021LatticeQuantumElectrodynamics, Cataldi2024Simulating2+1DSU2, Magnifico2025TensorNetworksLattice}. 
Their loopless hierarchical structure allows efficient contraction and polynomial computational cost, e.g., $\mathcal{O}(\Nsites d^2\chi^2 +\Nsites \chi^4)$ for binary TTNs \cite{Silvi2019TensorNetworksAnthology, Qian2022TreeTensorNetwork}.
This favorable scaling enables reaching relatively large bond dimensions compared to other higher-dimensional ansätze such as Projected Entangled Pair States (PEPS). 
However, the absence of internal loops means that TTNs do not automatically encode the area law in two or more dimensions \cite{Ferris2013AreaLawRealspace}. 
The precision of TTN simulations must therefore be carefully monitored, typically by studying convergence as $\chi$ increases.

Beyond ground-state optimization, TTNs can be used to simulate real-time dynamics \cite{Pavesic2025ConstrainedDynamicsConfinement}. 
A robust approach is the Time-Dependent Variational Principle (TDVP) \cite{Haegeman2011TimeDependentVariationalPrinciple, Bauernfeind2020TimeDependentVariational}. 
In TDVP, the time-dependent Schrödinger equation is projected onto the tangent space of the chosen TN manifold (e.g., TTNs of fixed $\chi$). 
This guarantees conservation of the norm and energy of the evolved state, with a computational scaling per time step that is comparable to a variational ground-state sweep \cite{Yang2020TimedependentVariationalPrinciple}.

The accuracy of TDVP depends on two key parameters: time step $\delta t$ and bond dimension $\chi$.
Choosing $\delta t$ too large compromises the approximation, while too small steps increase computational cost. 
Convergence must then be checked by systematically varying $\delta t$.
Similarly, during out-of-equilibrium evolution, entanglement generally grows, leading to a required $\chi$ big enough to maintain accuracy. \cite{Schuch2008EntropyScalingSimulability, Paeckel2019TimeevolutionMethodsMatrixproduct}.

In the simulations presented in the main text, we always initialize the dynamics from a product state and evolve it using a \emph{single-tensor} TDVP update at each step. 
To avoid variational locking and allow the entanglement support to grow when needed, we employed a pad of the initial tensor network up to $\chi$ (cf. single-tensor expansion ideas in \cite{Hubig2015StrictlySinglesiteDMRG, Gleis2023ControlledBondExpansion, Pavesic2025ConstrainedDynamicsConfinement}). 
This makes the simulation much faster on GPUs, since we do not need to use SVDs \cite{Krinitsin2025TimeEvolutionQuantum}.
Both the time step and the bond dimension adopted in our simulations are chosen sufficiently small and large, respectively, to ensure convergence of all reported observables. 
As shown in \cref{fig_convergence}, entanglement entropy and local observables converge well already at $\chi\!=\!150$ and $\delta t\!=\!10^{-2}$.
Remarkably, although TN dynamics are generally constrained to moderate times, we reached long enough evolution times on lattices up to $\Lambda = 8\times 8$ to unambiguously observe string breaking and related phenomena.
This is facilitated by the fact that, in the coupling regimes investigated, the dynamics remain effectively confined to the patch between the two static background charges, which keeps entanglement growth under control and enables reliable long-time simulations.

\section{Quantum fluctuations on observables}
\begin{figure}
\includegraphics[width=1\columnwidth]{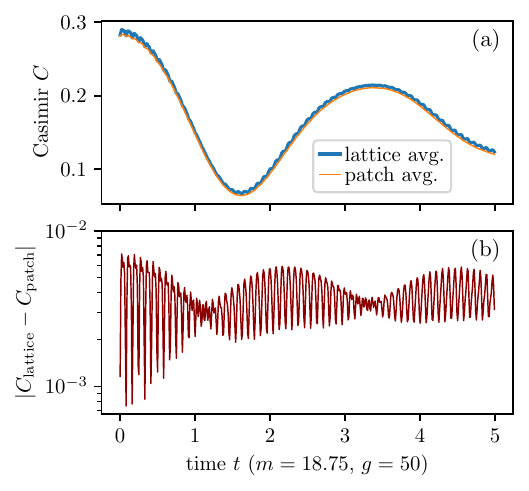}
\caption{\textbf{Fluctuations outside the patch.} Size of perturbative fluctuations outside the patch defined by the two static charges. 
Results are obtained from TTN simulation of a $\Lambda=8\times8$ lattice with a $4\times 4$ patch.}
\label{fig_fluctuations}
\end{figure}

Although the entire string dynamics remains confined within the patch, outside the patch, the system detects small perturbative fluctuations (due to hopping and magnetic terms in the \cref{eq_H}) which add noise to the observables averaged over the entire lattice. 
As shown in \cref{fig_fluctuations}, these fluctuations ($<10^{-2}$) can be clearly detected and removed from local observables, simply by comparing the single-site average over the patch and the corresponding one performed on the whole lattice (and then normalized on the patch, as in \cref{eq_local_observables}).
Conversely, fidelities cannot be cleaned up, as they are obtained by states intrinsically defined on the entire lattice.

\section{Detection of String Breaking via Casimir Minimum}
\begin{figure}
    \centering
    \includegraphics[width=1\columnwidth]{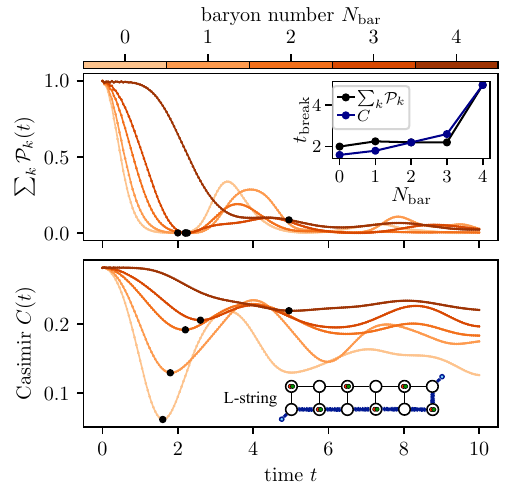}
    \caption{\textbf{Baryon Blockade mechanism with ED simulations.} Simulation of the baryon blockade slowdown obtained with ED on a $6\times 2$ patch in terms of the string manifold weight and the Casimir operator.}
    \label{fig4ED}
    
\end{figure}

The most direct way to detect string breaking is to evaluate the string-manifold weight $\mathcal{P}(t)$, summing the fidelities between the time-evolved state $\psi(t)$ and all the unbroken string states:
\begin{equation}
    \mathcal{P}(t) = \sum_k \mathcal{P}_k(t) = \sum_k \mathcal |\braket{\psi(t)|\psi_k}|^2.
\end{equation}
The minimum of $\mathcal{P}(t)$ signals maximal string breaking. 
However, measuring $\mathcal{P}(t)$ can be numerically expensive, as the number of unbroken string states grows fast with the size of the patch. 
For instance, the initial snake $S$-string, such as the one adopted to study the baryon blockade mechanism in \cref{fig4_baryon_blockade}, can resonate, through the effect of the plaquette term, with 203 different unbroken string states. 
Computing all these fidelities at each time step dramatically slows down the simulations. 
However, since string breaking reduces the energy related to the Casimir $C(t)$ by creating particles (as explained in the main text), we can alternatively use the minimum of $C(t)$ as a proxy for string breaking. 
To further justify this approach, we show that $\mathcal{P}(t)$ and $C(t)$ undergo a similar baryon blockade slowdown on a $6\times2$ lattice with ED simulations. 
As shown in \cref{fig4ED}, there is a strong correlation between the breaking time $t_{\text{break}}$, determined via $C(t)$ minimum, and the corresponding one obtained with the $\mathcal{P}(t)$ minimum. 
Moreover, this approach is free of the quantum fluctuation discussed above, facilitating the identification of the breaking point.
\end{document}